\documentclass[12pt,english]{article}
\usepackage[T1]{fontenc}
\usepackage[latin9]{inputenc}
\usepackage{geometry}
\usepackage{amsmath}
\usepackage{amssymb}

\makeatletter
\newenvironment{lyxlist}[1]
{\begin{list}{}
{\settowidth{\labelwidth}{#1}
 \setlength{\leftmargin}{\labelwidth}
 \addtolength{\leftmargin}{\labelsep}
 }}
{\end{list}}

\pdfoutput=1

\usepackage{babel}

\makeatother

\usepackage{babel}
\begin{document}

\title{\date{}\interfootnotelinepenalty=10000 A Relational Concept of Machian
Relativity}

\author{Herman Telkamp%
\thanks{Jan van Beverwijckstr. 104, 5017 JA Tilburg, Netherlands; email herman\_telkamp@hotmail.com%
}}
\maketitle
\begin{abstract}
Mach's principle fits into the wider \textquotedbl{}relational principle\textquotedbl{},
advocating that not only inertia, but also space and time emerge from
the interaction of matter. Concepts of a Machian/relational theory
are proposed, where inertia and energy are defined as mutual properties
between pairs of objects. Due to Berkeley, only radial motion represents
kinetic energy between (point) masses, which is the basis of anisotropic
inertia, which in turn underlies the relational principle. The Newtonian
definition of potential energy is considered a model for Machian inertia,
leading to a frame independent definition of Machian kinetic energy,
which comprises of the Newtonian terms (relative to the \textquotedbl{}fixed
stars\textquotedbl{}) and small anisotropic Machian energy terms between
objects. The latter account for relativistic trajectories, such as
the anomalous perihelion precession and Lense-Thirring frame dragging.
However, relativistic effects of remote observation (e.g. time dilation)
demand an isotropic model. A relational spacetime metric is derived,
which provides an isotropic coordinate transform of the anisotropic
Machian model, yielding a relational model which matches GR expressions
for relativistic trajectories and effects of remote observation. Therefore,
the experimental verification of GR in these cases holds automatically
for the relational model. The relational model fits the relational
principle (including Mach's principle) and it is argued that it includes
GR as a special case. The relational metric provides both contraction
and (unbound) expansion as a function of relative potential, i.e.
without invoking dark energy.\medskip{}

\end{abstract}

\section{Introduction\label{sec:Introduction}}

By \textit{Mach's principle} the inertia of a mass can only emerge
from the presence of other matter (planets, stars,\ldots{}). This
tells us mass inertia is a relative quantity, just like the force
of gravity is, and not an absolute property of an object. It is well
known that Einstein attempted for a long time to incorporate Mach's
principle into his general theory of relativity (GR), but finally
rejected the idea. This is remarkable, even though GR is extremely
successful in describing relativistic phenomena by modifying space
and time, while leaving inertia invariant. This, however, does not
necessarily imply that Einstein's spacetime concept is the only perceivable
model of relativity. After all, Mach's argument remains valid and
several authors demonstrated the inherently relativistic aspects of
the Machian principle. For instance, Sciama \cite{Sciama (1953)}
and recently Gogberashvili and Kanatchikov \cite{Gogberashvili & Kanatchikov (2011)}
explored relativistic cosmological implications. Others, e.g. Hofman
\cite{Hofman (1904)}, Schrödinger \cite{Schroedinger (1926)}, Barbour
and Bertotti \cite{Barbour Bertotti (1977)} considered particle models,
as we will do in this paper.

The approach taken here relies on GR, in particular on the Schwarzschild
solution, but mainly as a reference. Rather than a geometric field
theory, we follow the views of Bishop Berkeley regarding the anisotropy
of inertia (section \ref{sec:Mach's-Princ}) and extend the approach
taken by Schrödinger \cite{Schroedinger (1926)} into a fully relational
Machian framework (sections \ref{sec:Isotropic-v.s.-anisotropic},\ref{sec:Anisotropic Machian inertia},\ref{sec:Models-of-Machian}),
which enables straightforward calculation of relativistic trajectories,
such as the anomalous perihelion precession and Lense-Thirring frame-dragging
(section \ref{sec:Machian-relativity-vs.}).

One aspect that is absent in typical particle models is that of remote
observation, which is elementary to GR and which we will be adding
(section \ref{sec:The-relation-model}) in a generalized form to the
Machian model, by which the model also satisfies the wider relational
principle. This \textit{relational model}, i.e. the Machian model
subject to a relational metric, is derived from - so is consistent
with - the Schwarzschild solution. It therefore matches relativistic
effects of remote observation, such as gravitational time dilation
and redshift. The model can be applied to arbitrary configurations
of N-body systems, i.e. is not limited to specific GR solutions like
the Schwarzschild and Kerr metric.

Even though GR is used as a reference throughout this text, the present
relational model differs conceptually from Einstein's theory. In GR,
inertia and energy are attributed to an object, while these are mutual
properties between objects in the present approach. A striking difference
is that the Schwarzschild solution does not appear as a vacuum solution
in the Machian equivalent. The \textquotedbl{}vacuum\textquotedbl{}
solution seems to assume a background of a flat potential equal to
$\varphi_{o}$, the potential in our part of the universe (section
\ref{sub:Isotropic-model}). This is why the Schwarzschild solution
appears as a special case of the Machian representation, where the
latter applies to any level of background potential. The absence of
a true vacuum also explains why GR (without later modifications such
as dark energy) does not predict an accelerated expansion of the universe
\cite{Myrzakulov et al (2011)}: against a non-zero flat background
potential, particles do not completely lose their inertia when receding
from a mass kernel, while proper time and length approach asymptotic
coordinate time and length. In the relational model, on the contrary,
acceleration is explained by vanishing inertia and expanding spacetime
(section \ref{sub:Cosmological-implications}).

\section{Descartes, Berkeley, Mach and the relational model\label{sec:Mach's-Princ}}

Mach's principle fits into a long tradition of \textquotedbl{}relational
physics\textquotedbl{}, dating back to Descartes \cite{Descartes (1644)}
or even Aristotle. The relationalists propose, loosely stated, that
physically meaningful quantities only emerge from the interaction
of matter. This \textquotedbl{}relational principle\textquotedbl{}
not only concerns inertia (Mach's principle), but also space and time
need matter to exist. Consequently, a true vacuum does not exist,
even though there is emptiness. One way to capture this is by considering
matter not as the material object we see, but rather as the (infinite)
gravitational field it causes, i.e., the field is the matter.

The relational principle is philosophically motivated and it has proven
difficult to turn it into a consistent physical theory. In the present
approach, however, we pursue this relational view. It has strong implications.
Firstly, laws of nature must be independent of the chosen frame of
reference. This requirement is met in GR; the field equations hold
in any frame. Secondly, relational parameters (inertia, distance,
time) must dilute completely (i.e. vanish) while moving into empty
space; they gradually lose their physical meaning%
\footnote{Vanishing of length and time is to be understood as infinite expansion
of unit length and infinite increase of clock rate.%
}. In GR, inertia is invariant, yet space and time do dilute. Though,
only to a minor degree and certainly not completely in empty space;
in GR-vacuum, spacetime is (asymptotically) flat Minkowskian. It is
argued in sections \ref{sub:Isotropic-model} and \ref{sec:The-relation-model}
that the GR-vacuum is not a true vacuum and that GR represents a special
case of relational physics.

Berkeley's \cite{Berkeley (1721)} criticism of Newton's concept of
absolute space regards the notion that position or motion of a sole
object (point mass $m_{1}$) in otherwise empty space is unobservable;
there is no other physical object to relate the position of $m_{1}$
to. Any definition of position, velocity or orientation of $m_{1}$
is arbitrary and has no physical meaning, i.e. in the relational view
these quantities are physically inexistent in a one-body universe.
The same applies to the kinetic energy of the single object. So, how
could this object exhibit mass inertia? In agreement with Mach's principle,
it can not. This changes when a second (point) mass $m_{2}$ appears.
Due to the force of gravity, the two bodies will accelerate towards
each other and build up kinetic energy. Therefore, according to Mach,
$m_{1}$ and $m_{2}$ must have acquired some inertia due to each
others presence. In Berkeley's view, contrary to Newton, their radial
distance (separation) is the only meaningful geometrical parameter
among the two objects. Indeed, the orientation of the two-body system
in empty space is unobservable and, as pointed out by Berkeley, any
circular motion of the two bodies around each other is physically
meaningless (our frame of reference could as well be rotating in the
opposite direction). Therefore, by this \textquotedbl{}anisotropic
principle\textquotedbl{} of Berkeley, contrary to radial motion, tangential
motion has no inertia and does not represent kinetic energy between
the two bodies%
\footnote{Radial motion must be taken quite literally though for finite-size
objects: spin of either one of the bodies at constant separation implies
radial motion of mass elements of one body relative to the other body.
Therefore, spin \textit{does} represent kinetic energy between the
bodies.%
}. 

\medskip{}

\noindent Hence, from a Berkeleyan/Machian point of view, the Newtonian
kinetic energy attributed to the circular orbit of two bodies in empty
space is all virtual. Therefore, the 'real' (Machian) kinetic energy
of the revolving system ($m_{1}$,$m_{2}$), denoted $T_{12}$, is
zero. Machian kinetic energy may be interpreted as the part of the
Newtonian kinetic energy that would be dissipated in an inelastic
collision. Indeed, \textquotedbl{}freezing\textquotedbl{} the two
objects together stops any relative radial motion, including spin
of the bodies, but it does not affect the rotation or translation
of the total system in empty space. These latter motions are artificial,
unobservable, therefore this part of Newtonian kinetic energy is virtual.
The above picture changes if we would move $m_{1}$ and $m_{2}$ from
empty space into our universe, which we may represent by a hollow
sphere of mass $m_{o}$. Then, the same circular orbit of $m_{1}$
and $m_{2}$, implies (components of) radial motion of $m_{1}$ and
$m_{2}$ relative to $m_{o}$, as explained further in section \ref{sec:Models-of-Machian}.
Thus, the orbit involves non-zero Machian kinetic energies $T_{01}$
and $T_{02}$ of the subsystems ($m_{1}$,$m_{o}$) and ($m_{2}$,$m_{o}$),
respectively. (As pointed out in section \ref{sec:Models-of-Machian},
these two terms actually represent the Newtonian kinetic energies
of $m_{1}$ and $m_{2}$, defined in a CM frame attached to the \textquotedbl{}fixed
stars\textquotedbl{}). The Machian energy $T_{12}$ associated with
the subsystem ($m_{1}$,$m_{2}$) remains zero. If, however, $m_{1}$
and $m_{2}$ were spinning or in a non-circular orbit, this would
involve radial motion between these bodies, therefore $T_{12}>0$.\medskip{}

\noindent Berkeley's anisotropic principle thus entails the concept
of both inertia and kinetic energy as a mutual property between objects.
This concept is consistent with the epistemological requirement of
a frame independent definition of kinetic energy, as employed by Schrödinger
\cite{Schroedinger (1926)}. Expressions for Machian kinetic energy
will be provided in section \ref{sec:Models-of-Machian}. Notably,
Berkeley's principle also provides a rationale for the emergence of
space and time from the distribution of matter: along with radial
inertia, both radial distance and a certain concept of time (radial
acceleration) emerge from the advent of the second body in empty space,
while the tangential dimensions remain unobservable (i.e. inexistent)
until other bodies appear in those directions. However, from an increasing
distance, this collection of bodies gradually begins to appear as
one single body, finally shrinking to a point mass in otherwise empty
space, making space and time gradually dilute at larger scales and
ultimately vanish at infinity. Thus, Berkeley's principle underlies
the relational principle, which includes Mach's principle.

\section{Isotropic v.s. anisotropic model\label{sec:Isotropic-v.s.-anisotropic}}

As a common denominator in typical Machian particle models, the inertia
$\mu$ of an object of mass $m$ at position $r$ is related to the
gravitational potential of the other masses of the universe. Essentially
two approaches can be distinguished: the isotropic model and the anisotropic
model.

\subsection{Isotropic model\label{sub:Isotropic-model}}

The isotropic model is not consistent with Berkeley's anisotropic
principle. It, however, does match relativistic remote observation
phenomena as follows. Consider a mass $m$ having an isotropic inertia
defined proportional to local potential
\begin{equation}
\mu(r)=\frac{\varphi(r)}{\varphi_{o}}m\label{eq:mu iso}
\end{equation}
where $\varphi(r)$ is the total potential at $r$ due to all other
masses. The scaling factor $\varphi_{o}$ is equal to the background
potential in our part of the universe and is used for convenience;
it makes $\mu(x)=m$ wherever $\varphi(x)=\varphi_{o}$. This isotropic
inertia can straightforwardly explain gravitational time dilation
and gravitational redshift as follows.\medskip{}

\noindent A clock driven by a harmonic oscillator of mass $m$ and
unity spring constant is positioned at a distance $r$ from the center
of a spherical mass $M$ (planet), which exerts a potential $\varphi_{{\scriptscriptstyle M}}(r)$
at the position of $m$. Assume a background potential $\varphi_{o}$,
then the isotropic inertia of mass $m$ is
\begin{equation}
\mu(r)=\frac{\varphi_{o}+\varphi_{{\scriptscriptstyle M}}(r)}{\varphi_{o}}m.
\end{equation}
The differential equation of the oscillator is $x=-\mu(r)\ddot{x}$,
having solution: $x(t)=a\sin(\omega t+\phi_{o})$. Hence,
\begin{equation}
\mu(r)=-\frac{x}{\ddot{x}}=\frac{1}{\omega^{2}(r)}.\label{eq:mu osc-1}
\end{equation}
Elapsed clock time $\tau_{{\scriptscriptstyle R}}(r)$ between two
events is proportional to clock frequency $\omega(r)$. If we have
two identical clocks at arbitrary positions $r_{A}$ and $r_{B}$,
then we may compare the elapsed times of the clocks by the \textit{relational
dilation factor }
\begin{equation}
\alpha_{{\scriptscriptstyle R}}=\alpha_{{\scriptscriptstyle R}}(r_{A},r_{B})=\frac{\tau_{{\scriptscriptstyle R}}^{2}(r_{B})}{\tau_{{\scriptscriptstyle R}}^{2}(r_{A})}=\frac{\omega^{2}(r_{B})}{\omega^{2}(r_{A})}=\frac{\mu(r_{A})}{\mu(r_{B})}=\frac{\varphi(r_{A})}{\varphi(r_{B})}\label{eq:Mach time dilation}
\end{equation}
Following (\ref{eq:Mach time dilation}), the ratio of elapsed times
appears inversely proportional to the ratio of the corresponding potentials.
In the special case where we compare the clock at $r_{B}=r$ with
a clock at infinity $r_{A}=\infty$, we have the specific relational
dilation factor
\begin{equation}
\hat{\alpha}_{{\scriptscriptstyle R}}=\alpha_{{\scriptscriptstyle R}}(\infty,r)=\frac{\tau_{{\scriptscriptstyle R}}^{2}(r)}{\tau_{{\scriptscriptstyle R}}^{2}(\infty)}=\frac{\varphi(\infty)}{\varphi(r)}=\frac{\varphi_{o}}{\varphi_{o}+\varphi_{{\scriptscriptstyle M}}(r)}\label{eq:Mach time dilation-1}
\end{equation}
In GR, according to the Schwarzschild metric, time dilation relates
\textit{local time} (i.e. stationary proper time) $\tau_{{\scriptscriptstyle S}}(r)$
to \textit{coordinate time} $t=\tau_{{\scriptscriptstyle S}}(\infty)$
as
\begin{equation}
\alpha_{{\scriptstyle {\scriptscriptstyle S}}}=1{\scriptstyle -}\frac{r_{s}}{r}=\frac{\tau_{{\scriptscriptstyle S}}^{2}(r)}{\tau_{{\scriptscriptstyle S}}^{2}(\infty)}\label{eq:GR time dila}
\end{equation}
where the Schwarzschild radius $r_{s}=2GM/c^{2}$. The specific relational
dilation factor (\ref{eq:Mach time dilation-1}) and the Schwarzschild
dilation factor (\ref{eq:GR time dila}) become virtually identical
if we assume
\begin{equation}
\varphi_{o}={\scriptstyle -\frac{1}{2}}c^{2}\label{eq:phi o}
\end{equation}
since then $r_{s}/r=\varphi_{{\scriptscriptstyle M}}(r)/\varphi_{o}$
and substituting this in (\ref{eq:Mach time dilation}) yields
\begin{equation}
\frac{\tau_{{\scriptscriptstyle R}}^{2}(r)}{\tau_{{\scriptscriptstyle R}}^{2}(\infty)}=\hat{\alpha}_{{\scriptscriptstyle R}}=\frac{1}{1+\frac{r_{s}}{r}}\approx1{\scriptstyle -}\frac{r_{s}}{r}=\alpha_{{\scriptstyle {\scriptscriptstyle S}}}=\frac{\tau_{{\scriptscriptstyle S}}^{2}(r)}{\tau_{{\scriptscriptstyle S}}^{2}(\infty)}.\label{eq:matching}
\end{equation}
Hence, the time dilation predicted by the isotropic Machian model
is virtually identical%
\footnote{Above the surface of earth, the ratio $r_{s}/r$ is extremely small
($<10^{-5}$) for any $r$. Therefore, the relative error in approximation
(\ref{eq:matching}) is negligible ($<10^{-10}$). %
} to GR time dilation. This has a remarkable consequence: from a Machian
point of view, the vacuum of the Schwarzschild solution doesn't look
empty really, rather implicitly assumes a flat background potential
$\varphi_{o}$, which was explicitly assumed in the Machian model.
Recall that the Schwarzschild metric is asymptotically Minkowskian
such that Newtonian absolute spacetime is recovered at infinity. Hence,
gravity appears in GR as a local curvature of a flat spacetime. The
Machian model, on the other hand, is purely relational; $\alpha_{{\scriptscriptstyle R}}$
is a dilation parameter between two arbitrary potentials at locations
$r_{A}$ and $r_{B}$. Contrary to GR, clock rate is unbound for a
clock moving into empty space, which observation is in agreement with
the relational principle. GR's concept of absolute \textquotedbl{}coordinate
time\textquotedbl{} (as well as \textquotedbl{}coordinate distance\textquotedbl{})
at infinity in empty space is therefore considered non-Machian. This
has been noted already by several authors, e.g. \cite{Lichtenegger (2011)}.
However, as pointed out above, assuming a flat background potential
$\varphi_{o}$ in the GR \textquotedbl{}vacuum\textquotedbl{} turns
the relational metric into an (asymptotically) flat Minkowski metric,
thus giving rise to the thought that GR represents a special case
of the relational metric. Further support of this argument is given
in section \ref{sub:Anomalous-precession-of} on the anomalous perihelion
precession. The unbound relational metric is formalized in the relational
model of section \ref{sub:General-form-of}.\medskip{}

\noindent Like in GR, gravitational redshift is just a different manifestation
of gravitational time dilation. Indeed, (\ref{eq:Mach time dilation})
provides the Machian ratio between emitted and observed frequency,
according to
\begin{equation}
\alpha_{{\scriptstyle {\scriptscriptstyle R}}}(r_{obs},r_{emit})=\frac{\nu_{emit}^{2}}{\nu_{obs}^{2}}=\frac{\varphi(r_{obs})}{\varphi(r_{emit)}}
\end{equation}
which in the Schwarzschild case for a source at distance $r$ from
$M$ and an observer at infinity (potential $\varphi_{o}$) becomes
\begin{equation}
\frac{\nu_{emit}^{2}(r)}{\nu_{obs}^{2}(\infty)}=\hat{\alpha}_{{\scriptscriptstyle R}}=\frac{\varphi(\infty)}{\varphi(r_{emit)}}=\frac{\varphi_{o}}{\varphi_{o}+\varphi_{{\scriptscriptstyle M}}(r)}\approx\alpha_{{\scriptstyle {\scriptscriptstyle S}}},
\end{equation}
consistent with gravitational redshift in the Schwarzschild spacetime.

\subsection{Anisotropic model\label{sub:Anisotropic-model}}

As an example we consider the inertia of mass $m$ relative to a second
mass $M$ at a separation $r$. Due to Berkeley, both inertia and
the associated kinetic energy exclusively concern \textit{radial motion}
of $m$, i.e. motion in the direction of $M$, making inertia anisotropic.
Schrödinger \cite{Schroedinger (1926)} used an anisotropic model
to derive, without invoking the spacetime concept, the well known
GR expression for the anomalous perihelion precession of planetary
orbits. The key to this result in the paper of Schrödinger is the
small additional inertia of the planet into the direction of the sun.
Thus, it is precisely the anisotropy of inertia that accounted for
this convincing result. Stated otherwise, isotropic models can not
explain the anomalous precession (as pointed out in Appendix A). Even
so, there seems to exist a common understanding that anisotropic models
are ruled out by the famous experiments by Hughes (1960) and Drever
(1961) \cite{Drever (1961)}, which yielded extremely tight upper-bounds
on the anisotropy of nuclear magnetic resonance%
\footnote{Anisotropy may not be observable in resonance frequencies, it does
manifest itself, though, in the exchange of radial and tangential
kinetic energy, as in an elliptic orbit. In fact, the anomalous precession
may be considered as experimental evidence of the anisotropic universe.%
}, thus indicating inertial isotropy of the universe. Indeed, the anisotropic
model (like the one used by Schrödinger) applied to the harmonic oscillator
of section \ref{sub:Isotropic-model} shows different frequencies
for oscillations in the radial and tangential direction: time dilation
by this model \textit{only} occurs in the radial direction of oscillation,
in disagreement with the Hughes-Drever experiments and in disagreement
with GR. As a consequence we conclude, even without the Hughes-Drever
experiments, that the anisotropic model fails to explain (the isotropic)
gravitational time dilation and redshift.

\subsection{A way out\label{sub:a-Way-out}}

So a Machian theory has to deal with seemingly contradicting requirements.
It must be isotropic and anisotropic at the same time, while anisotropy
is considered conflicting with Hughes-Drever experiments. This is
essentially where the Machian doctrine got stuck, while GR passed
all the tests. Note that GR's geometry involves \textit{two} elementary
concepts: time dilation and length contraction, where the first is
isotropic and the latter is anisotropic, while the typical Machian
approach only employs a single concept, i.e. variable inertia, either
isotropic or anisotropic. Apparently, the Machian concept of variable
inertia must be enhanced with a second concept to become GR compatible.
Nevertheless, as shown in section \ref{sec:Machian-relativity-vs.},
anisotropic inertia alone can correctly model the relativistic trajectory
of a system, like the anomalous precession (section \ref{sub:Anomalous-precession-of})
and frame dragging (section \ref{sub:Frame-dragging}). Sections \ref{sec:Anisotropic Machian inertia}
and \ref{sec:Models-of-Machian} provide the elementary expressions
for such a model. It is shown in section \ref{sec:The-relation-model}
that observational effects, like gravitational time dilation and gravitational
red shift, may be accounted for by an isotropic coordinate transform
of the anisotropic model.

\section{Anisotropic Machian inertia and energy\label{sec:Anisotropic Machian inertia}}

We briefly review the Newtonian definition of potential energy from
a relational perspective and consider the problematic Newtonian definition
of kinetic energy. From there we arrive at definitions of Machian
inertia and energy.

\subsection{Machian potential energy = Newtonian potential energy}

A single object $m_{i}$ at an arbitrary position $r_{i}$ in empty
space has zero potential energy: $V_{i}=m_{i}\varphi(r_{i})=0$. Indeed,
the gravitational potential $\varphi$ is nil everywhere in empty
space. A second object, $m_{j}$, is required to obtain a non-zero
potential energy $V_{ij}$ of the system ($m_{i}$,$m_{j}$). Let
$\varphi_{j}(r_{i})$ denote the gravitational potential due to $m_{j}$
at the position of $m_{i}$, then
\begin{equation}
V_{ij}=m_{i}\varphi_{j}(r_{i})=m_{j}\varphi_{i}(r_{j})=V_{ji}\label{eq:V 1 2}
\end{equation}
which equals the energy required to bring $m_{i}$ from infinity to
a distance $r_{ij}$ from $m_{j}$. Or, equivalently, the energy to
bring $m_{j}$ from infinity to a distance $r_{ij}$ from $m_{i}$.
Thus, the potential energy $V_{ij}$ belongs as much to $m_{i}$ as
to $m_{j}$. It doesn't exist without the other mass present, which
is a relational viewpoint. Hence, potential energy is a property of
the system ($m_{i}$,$m_{j}$)\textit{.} In general, for a N-body
system, we obtain the total potential energy by adding up for all
possible pairs (while avoiding double counting, since $V_{ij}=V_{ji}$)
\begin{equation}
V=\underset{{\scriptstyle i}}{{\scriptstyle {\textstyle \sum}}}\underset{{\scriptstyle j>i}}{{\textstyle \sum}}\, V_{ij}\qquad{\scriptstyle i,j=1..N}.\label{eq:V}
\end{equation}
In conclusion, Newtonian potential energy is a mutual property between
every pair of objects and its value only varies with the separation
between the objects, therefore is frame independent and fits the relational
view.

\subsection{Machian inertia and Machian kinetic energy}

\subsubsection*{Newtonian kinetic energy}

Contrary to potential energy, classical kinetic energy is a property
of an object $m_{i}$ and its value $T_{i}={\scriptscriptstyle \frac{1}{2}}m_{i}v_{i}^{{\scriptscriptstyle 2}}$
is dependent on the choice of frame, while the value of the inertia
parameter $m_{i}$ is not related to the presence of other mass. So
it expresses the absoluteness of space and inertia and it clearly
does not reflect Mach's views. The problem of absolute space and inertia
becomes most apparent in the classical energy balance of an isolated
system ($E$ represents the conserved total energy)
\begin{equation}
E=T+V\label{eq:Etotal}
\end{equation}
where $E$ and $V$ are frame independent, while $T$ ($=\!\!\Sigma\, T_{i}$)
isn't. Obviously, one can not create or destroy energy by moving a
frame, which makes the frame dependency of $T$ unphysical. So, a
consistent formulation of energy conservation requires kinetic energy
to be defined independently of the frame too. Moreover, (\ref{eq:Etotal})
strongly suggests kinetic energy, more specifically, the inertia parameter,
must be a reflection of potential energy to provide this consistency.

\subsubsection*{Machian inertia}

We let the potential energy $V_{ij}$ between two bodies be a model
of their mutual Machian inertia $\mu_{ij}$
\begin{equation}
V_{ij}=\mu_{ij}\varphi_{o}\label{eq: Vij}
\end{equation}
i.e., inertia \textit{is} potential energy (the scaling constant is
conveniently chosen to be $\varphi_{o}$, the background potential
in our part of the universe). Accordingly, we define the \textit{Machian
partial inertia} between objects $m_{i}$ and $m_{j}$ as
\begin{equation}
\mu_{ij}=\frac{V_{ij}}{\varphi_{o}}=\frac{m_{i}\varphi_{j}(r_{i})}{\varphi_{o}}=\frac{m_{j}\varphi_{i}(r_{j})}{\varphi_{o}}=\mu_{ji}.\label{eq:mu ij}
\end{equation}
By this definition, inertia is a property of the pair ($m{}_{i}$,$m{}_{j}$).
So multiple partial inertiae are associated with a single object,
exactly like all the partial potential energies associated with the
object. Definition (\ref{eq:mu ij}) is mutual and symmetrical, which
implies that two bodies of totally different mass (say the sun and
a neutron) exert an equal inertia to one and another, just like they
exert an equal force of gravity to one and another. A consequence
of the relational nature of inertia is - contrary to Newton's universe
- that in the Machian model we have to account explicitly for the
inertia caused by all the masses of the universe. To do so, we adopt
the hollow sphere model of the universe, also used by Einstein and
Schrödinger, by which we may treat the masses of the universe as a
single object, $m_{o}$. The potential inside of the hollow sphere
is flat and equal to $\varphi{}_{o}$, the background potential in
our part of the universe. Therefore, the inertia $\mu_{oi}$ between
an arbitrary object $m{}_{i}$ and the universe is simply $\mu_{oi}=m_{i}$.
Note that the inertia $\mu_{ij}$ between two arbitrary objects is,
in general, negligible relative to $\mu_{oi}=m{}_{i}$ and $\mu_{oj}=m{}_{j}$,
because $\varphi{}_{o}$, the potential due to the universal masses,
is extremely large.

\subsubsection*{Machian kinetic energy between point particles}

Using the above definition of partial inertia (\ref{eq:mu ij}) and
recalling Berkeley's argument that only radial motion between (point)
masses represents kinetic energy, we arrive at the following expression
of Machian kinetic energy between two point participles $m_{i}$ and
$m_{j}$
\begin{equation}
T_{ij}={\scriptstyle \frac{1}{2}}\mu_{ij}\dot{r}_{ij}^{2}.\label{eq:T ij}
\end{equation}
The Machian equation of motion for two masses falling to each other
in empty space (see appendix B) reads
\begin{equation}
\ddot{r}_{ij}=\frac{\varphi_{o}}{r_{ij}}({\scriptstyle {\textstyle 1}}+\frac{\dot{r}_{ij}^{2}}{{\textstyle {\textstyle {\textstyle {\scriptstyle 2}}}}\varphi_{o}}).\label{eq:kinematic eq empty-1}
\end{equation}
This differential equation lets $m_{i}$ and $m_{j}$ accelerate towards
each other until $\ddot{r}_{ij}=0$, at which point the relative velocity
$|\dot{r}_{ij}|$ attains its maximum value
\begin{equation}
|\dot{r}_{ij}|_{max}=\sqrt{{\scriptstyle -}2\varphi_{o}}.\label{eq:speed limit-1}
\end{equation}
According to (\ref{eq:phi o}), $\varphi_{o}={\scriptstyle -}{\scriptstyle \frac{1}{2}}c^{2}$,
hence, $|\dot{r}_{ij}|_{max}=c$, the speed of light. This is not
a surprise, as we recognize the Lorentz factor $\gamma$ in Eq.(\ref{eq:kinematic eq empty-1}):
\begin{equation}
({\scriptstyle {\textstyle 1}}+\frac{\dot{r}_{ij}^{2}}{{\textstyle {\textstyle {\textstyle {\scriptstyle 2}}}}\varphi_{o}})^{{\scriptscriptstyle -1/2}}=({\scriptstyle {\textstyle 1}}{\scriptstyle -}\frac{\dot{r}_{ij}^{2}}{c^{2}})^{{\scriptscriptstyle -1/2}}=\gamma.\label{eq:Lorentz factor-1}
\end{equation}
Appendix B shows that the velocity limit $c$ also holds in a non-empty
space. Ergo, the speed of light is a consequence of the assumed Machian
inertia, i.e. is not a postulate in the Machian context.

\noindent \medskip{}

\noindent The total kinetic energy of a system of N point masses follows
by straightforward summation (\ref{eq:T ij}) over all different pairs
\begin{equation}
T=\underset{{\scriptstyle i}}{{\textstyle \sum}}\underset{{\scriptstyle j>i}}{{\textstyle \sum}}T_{ij}\qquad{\scriptstyle i,j=1,..,N}
\end{equation}
The relationship between Machian and Newtonian kinetic energy will
be clarified in the next section.

\section{Machian kinetic energy between finite size objects\label{sec:Models-of-Machian}}

Eq. (\ref{eq:T ij}) provides the elementary expression of kinetic
energy between two point particles. The kinetic energy between two
solid objects of arbitrary shape and mass distribution is essentially
an integral of the kinetic energy between infinitesimal small volume
elements of the two objects over their respective volumes. For Machian
calculations of celestial mechanics we need models of kinetic energy
between two massive spheres (both translating and spinning) and between
a massive sphere and the hollow sphere (representing the universe).
These models are provided in this section.

\subsection{Translational kinetic energy\label{sub:Translational-kinetic-energy}}

\subsubsection*{Between an object and the universe}

The homogeneity of the universe allows assuming a constant background
potential $\varphi_{o}$, at least on the scale of the solar system.
This makes the hollow sphere model ($m_{o}$) an accurate representation.
Constant potential implies that the partial inertia between an object
$m_{i}$ and the heavenly masses is constant, so does not depend on
the position, particular shape or mass density distribution of the
object. By definition (\ref{eq:mu ij}) we obtain $\mu_{oi}=m{}_{i}$,
the Newtonian value. With the hollow sphere in place, the universe
has become just another object, $m{}_{o}$. Motion of an object $m{}_{i}$
along any trajectory through space implies components of radial velocity
relative to the cosmic masses. The direction of the motion in the
universe does not matter for the amount of kinetic energy it represents.
Just the magnitude $v_{oi}$ of the velocity. Therefore, in the case
of a body $m{}_{i}$ moving through the universe, we have a translational
Machian kinetic energy $T_{v_{oi}}$ between $m_{i}$ and $m_{o}$
equal to the Newtonian kinetic energy $T_{i}$ of object $m_{i}$
\begin{equation}
T_{v_{oi}}={\scriptscriptstyle \frac{1}{2}}\mu_{oi}v_{oi}^{{\scriptscriptstyle 2}}={\scriptscriptstyle \frac{1}{2}}m_{i}v_{i}^{{\scriptscriptstyle 2}}=T_{i}\label{eq:Tv,io-1}
\end{equation}
provided that the \textquotedbl{}absolute\textquotedbl{} velocity
$v{}_{i}$ is defined in the CM frame attached to $m{}_{o}$, so that
$v{}_{i}=v_{oi}$.

\subsubsection*{Between two solid spheres}

By definition (\ref{eq:mu ij}), the partial inertia between the bodies
is $\mu_{ij}=V_{ij}/\varphi_{o}$. For conservation of energy, equal
potential energies for spheres and point particles implies equal kinetic
energies. Hence, \textit{translational} kinetic energy between solid
spheres is the same as between point particles (\ref{eq:T ij}), therefore
it only has a radial component
\begin{equation}
T_{v_{ij}}=T_{\dot{r}_{ij}}={\scriptscriptstyle \frac{1}{2}}\mu_{ij}\dot{r}_{ij}^{2}\qquad\quad{\scriptstyle i,j\neq0}\label{eq:T r ij}
\end{equation}
since the tangential component of motion does not represent kinetic
energy between the bodies, i.e.
\begin{equation}
T_{\dot{\phi}_{ij}}=T_{\dot{\theta}_{ij}}=0\qquad\quad{\scriptstyle i,j\neq0}
\end{equation}
where $\phi_{ij}$ and $\theta_{ij}$ denote the perpendicular tangential
directions in spherical coordinates relative to the CM of the two
bodies.

\subsection{Spin energy\label{sub:Spin-energy}}

Spin of body $m_{i}$ and/or $m_{j}$ implies Machian kinetic energy
of the subsystem ($m_{i}$,$m_{j}$), even if the spinning bodies
stay at constant radius. Arbitrary volume elements of these two bodies
do have a non-zero radial velocity component relative to each other,
as their distance varies during spin of any of the two bodies. Therefore,
spin of a body relative to another body, represents radial velocity
between masses, hence, Machian kinetic energy. The total kinetic energy
of the spinning bodies follows from integrating the Machian kinetic
energies between volume elements of the respective bodies over the
volumes of the bodies. Instead of solving the (rather involved) integrals,
one can derive the solutions as well heuristically by formulating
the Machian analogy of Newtonian expressions for spinning bodies,
which are based on the same integrals.

\subsubsection*{Spin energy relative to the universe}

Consider a solid sphere $m_{i}$ of radius $R_{i}$, spinning at an
angular velocity $\omega_{i}$ relative to the CM frame attached to
hollow sphere $m_{o}$. The Newtonian spin energy is
\begin{equation}
T_{\omega_{i}}={\scriptscriptstyle \frac{1}{5}}m_{i}R_{i}^{2}\omega_{i}^{2}\label{eq:T w i}
\end{equation}
In an inelastic collision of $m_{i}$ and $m_{o}$, the spin of $m_{i}$
will vanish completely. Hence, the Newtonian value is all Machian
energy, therefore the analogous Machian expression is simply
\begin{equation}
T_{\omega_{oi}}={\scriptscriptstyle \frac{1}{5}}\mu_{oi}R_{i}^{2}\omega_{i}^{2}={\scriptscriptstyle \frac{1}{5}}m_{i}R_{i}^{2}\omega_{i}^{2}\label{eq:T w io}
\end{equation}
If we add an object $m_{j}$, then we also get (non-Newtonian) Machian
terms between $m_{i}$ and $m_{j}$:

\subsubsection*{Spin energy between two solid spheres}

The case of two spinning spheres appears to be very Machian. One has
to bear in mind that circular orbit of the spheres at a rate $\dot{\phi}_{ij}$,
even when the spheres are glued to each other, introduces a spin of
both bodies relative to the frame of reference, while this spin clearly
does not represent spin energy between the two spheres. Using this
heuristic, we derive a frame independent expression for the spin energy
between two spheres $m{}_{i}$ and $m{}_{j}$. Assume the two spheres
are at separation $r_{ij}$ and have radii $R{}_{i}$ and $R{}_{j}$.
Imagine for a moment they take fixed positions in the frame of reference
($\dot{\phi}_{ij}=0$). Sphere $m_{i}$ has a spin $\omega_{i}$ relative
to the frame, while $m_{j}$ is at rest . In this case, the Machian
spin energy of $m_{i}$ relative to $m_{j}$ is obtained by generalization
of (\ref{eq:T w io})
\begin{equation}
T_{\omega_{i/j}}={\scriptscriptstyle \frac{1}{5}}\mu_{ij}R_{i}^{2}\omega_{i}^{2}.\label{eq:T w i/j}
\end{equation}
In fact, we have only replaced universe $m_{o}$ in (\ref{eq:T w io})
by object $m_{j}$. Obviously, for the opposite case, where $m_{j}$
is spinning and $m_{i}$ is at rest, we have
\begin{equation}
T_{\omega_{j/i}}={\scriptscriptstyle \frac{1}{5}}\mu_{ij}R_{j}^{2}\omega_{j}^{2}.\label{eq:T w j/i}
\end{equation}
If both objects are orbiting each other at a rate $\dot{\phi}_{ij}$,
we must subtract the contribution of this rotation from $\omega_{i}$
and $\omega_{j}$ to obtain a frame independent expression for the
spin energy of the system ($m_{i}$,$m_{j}$)
\begin{equation}
T_{\omega_{ij}}=T_{\omega_{i/j}}+T_{\omega_{j/i}}={\scriptscriptstyle \frac{1}{5}}\mu_{ij}R_{i}^{2}(\omega_{i}{\scriptstyle -}\dot{\phi}_{ij})^{2}+{\scriptscriptstyle \frac{1}{5}}\mu_{ij}R_{j}^{2}(\omega_{j}{\scriptstyle -}\dot{\phi}_{ij})^{2}.\label{eq:T w ij}
\end{equation}
So, spins must be considered relative to the common axis which connects
the centers of the two spheres. Indeed, the Machian spin energy between
$m_{i}$ and $m_{j}$ is completely gone if we would freeze the spheres
together onto this imaginary axis%
\footnote{In the general case, the objects may also spin in the perpendicular
direction $\theta_{ij}$, giving an additional spin energy $T_{\psi_{ij}}=T_{\psi_{i/j}}+T_{\psi_{j/i}}={\scriptscriptstyle \frac{1}{5}}\mu_{ij}R_{i}^{2}(\psi_{i}{\scriptstyle -}\dot{\theta}_{ij})^{2}+{\scriptscriptstyle \frac{1}{5}}\mu_{ij}R_{j}^{2}(\psi_{j}{\scriptstyle -}\dot{\theta}_{ij})^{2}.$
For simplicity, we will leave out this component hereafter, as it
can be avoided.%
}.

\noindent \medskip{}

\noindent Like with translational energy, since $\mu_{ij}$ is typically
extremely small, the spin energy between two spheres is negligible
in comparison with each sphere's (Newtonian) spin energy relative
to the universe. Yet, it does account for relativistic effects, such
as the Lense-Thirring precession, as will be pointed out in section
\ref{sub:Frame-dragging}.

\subsection{Example - Machian energy of two orbiting bodies \label{sub:Example two bodies}}

In universe \textit{$O$} (background potential $\varphi_{o}$), two
non-spinning spherical bodies, $m$ and $M$, are orbiting each other
at angular velocity $\dot{\phi}$ and radius $r=r_{m}+r_{{\scriptscriptstyle M}}$,
where $r_{m}$ and $r_{{\scriptscriptstyle M}}$ are the distances
from $m$ and $M$ to their CM. We assume $m\ll M$. The total energy
of the 3-body system ($m$,$M$,$O$) is assembled as follows
\[
E=T+V
\]
\[
T=T_{mO}+T_{MO}+T_{mM}
\]
\begin{equation}
V=V_{mO}+V_{MO}+V_{mM}=(m+M+\mu)\varphi_{o}\label{eq:V mercury-1}
\end{equation}
 where the kinetic energy components are (in polar coordinates)
\begin{equation}
T_{mO}={\scriptstyle \frac{1}{2}}m\dot{r}_{m}^{2}+{\scriptstyle \frac{1}{2}}mr_{m}^{2}\dot{\phi}^{2}\label{eq:Tmo merc-2}
\end{equation}
\begin{equation}
T_{MO}={\scriptstyle \frac{1}{2}}{\scriptstyle M}\dot{r}_{{\scriptscriptstyle M}}^{2}+{\scriptstyle \frac{1}{2}}{\scriptstyle M}r_{{\scriptscriptstyle M}}^{2}\dot{\phi}^{2}\label{eq:T Mo merc-1}
\end{equation}
\begin{equation}
T_{mM}={\scriptstyle \frac{1}{2}}\mu\dot{r}^{2}\label{eq:TmM merc-1}
\end{equation}
and where $\mu=\mu_{mM}=-GmM/\varphi_{o}r$. The radius $r_{m}=\frac{M}{M+m}r$
and $r_{{\scriptscriptstyle M}}=\frac{m}{M+m}r$. Considering $m\ll M$,
substitution into (\ref{eq:Tmo merc-2}) and (\ref{eq:T Mo merc-1})
gives
\begin{equation}
T_{mO}={\scriptstyle \frac{1}{2}}m\left(\frac{M}{M+m}\right)^{2}\dot{r}^{2}+{\scriptstyle \frac{1}{2}}m\left(\frac{M}{M+m}\right)^{2}r^{2}\dot{\phi}^{2}\approx{\scriptstyle \frac{1}{2}}m\dot{r}^{2}+{\scriptstyle \frac{1}{2}}mr^{2}\dot{\phi}^{2}\label{eq:Tmo merc-1-1}
\end{equation}
 and
\begin{equation}
T_{MO}={\scriptstyle \frac{1}{2}}M\left(\frac{m}{M+m}\right)^{2}\dot{r}^{2}+{\scriptstyle \frac{1}{2}}M\left(\frac{m}{M+m}\right)^{2}r^{2}\dot{\phi}^{2}
\end{equation}
So, $T_{MO}\thickapprox\frac{m}{M}T_{mO}\ll T_{mO}$, therefore $T_{MO}$
can be neglected. Accordingly, we eliminate the associated constant
potential energy $V_{MO}$ from the equation. In conclusion, we obtain
the following simplified expression for the orbital energy of $m$
relative to $M$ and $O$:
\begin{equation}
E=T_{mO}+T_{mM}+V_{mO}+V_{mM}={\scriptstyle \frac{1}{2}}(m+\mu)\dot{r}^{2}+{\scriptstyle \frac{1}{2}}mr^{2}\dot{\phi}^{2}+(m+\mu)\varphi_{o}\label{eq:E}
\end{equation}
If, in addition, the two bodies $m$ and $M$ spin at angular velocities
$\omega_{m}$ and $\omega_{M}$ relative to $O$, then we must add
spin energy
\begin{equation}
T_{\omega}=T_{\omega_{mO}}+T_{\omega_{MO}}+T_{\omega_{mM}}={\scriptscriptstyle \frac{1}{5}}mR_{m}^{2}\omega_{m}^{2}+{\scriptscriptstyle \frac{1}{5}}MR_{{\scriptscriptstyle M}}^{2}\omega_{{\scriptscriptstyle M}}^{2}\:+\:{\scriptscriptstyle \frac{1}{5}}\mu\left(R_{m}^{2}(\omega_{m}{\scriptstyle -}\dot{\phi})^{2}+R_{{\scriptscriptstyle M}}^{2}(\omega_{{\scriptscriptstyle M}}{\scriptstyle -}\dot{\phi})^{2}\right)\label{eq:T w}
\end{equation}
If $m$ is only an infinitesimal small test particle, then $R_{m}=0$,
which simplifies (\ref{eq:T w}) to
\begin{equation}
T_{\omega}={\scriptscriptstyle \frac{1}{5}}MR_{{\scriptscriptstyle M}}^{2}\omega_{{\scriptscriptstyle M}}^{2}\:+\:{\scriptscriptstyle \frac{1}{5}}\mu R_{{\scriptscriptstyle M}}^{2}(\omega_{{\scriptscriptstyle M}}{\scriptstyle -}\dot{\phi})^{2}\label{eq:T wM}
\end{equation}
The simplified equations (\ref{eq:E}) and (\ref{eq:T wM}) fit well
known celestial problems (see section \ref{sec:Machian-relativity-vs.}).

\section{Consistency of the Machian model and GR\label{sec:Machian-relativity-vs.}}

Using the Machian model, i.e. the definitions of anisotropic inertia
and Machian energy of section \ref{sec:Anisotropic Machian inertia}
and \ref{sec:Models-of-Machian}, we derive in this section expressions
for the anomalous perihelion precession and for Lense-Thirring frame-dragging,
both matching the well-known GR expressions. That is, as long as we
assume a background potential equal to $\varphi_{o}$. Note that the
Machian model does not involve the spacetime concept.

\subsection{Anomalous precession of Mercury's perihelion\label{sub:Anomalous-precession-of}}

\noindent The anomalous precession is a classical test case of GR.
Using the Schwarzschild solution%
\footnote{Remarkably, in the case of Mercury's precession, the Schwarzschild
solution has been successfully applied to a non-vacuum space of background
potential $\varphi_{o}$, which supports the earlier comment that
the Schwarzschild vacuum does not actually appear empty.%
}, one derives the following relativistic energy equation for a small
test particle $m$ orbiting a gravitating central mass $M$
\begin{equation}
{\scriptstyle \frac{1}{2}}m\dot{r}^{2}+{\scriptstyle \frac{1}{2}}\frac{mh^{2}}{r^{2}}-\frac{GmM}{r}-\frac{GmMh^{2}}{c^{2}r^{3}}=const.\label{eq:energy mercury}
\end{equation}
Here, $m$ represents Mercury's mass, $M$ is the solar mass, $G$
Newton's constant and $h=r^{2}\dot{\phi}$ is the (constant) momentum
of Mercury's orbit per unit mass, $r$ is the radius of Mercury's
orbit and $\dot{\phi}$ the orbital angular velocity. This expression
is Newtonian, except for the last term on the left, which is purely
GR and accounts completely for the anomalous precession. From (\ref{eq:energy mercury})
one can derive the famous expression of the anomalous precession per
revolution
\begin{equation}
\Delta\phi=\frac{6\pi G^{2}M^{2}}{c^{2}h^{2}}\label{eq:GR Anomalous prec}
\end{equation}
\medskip{}

\noindent Next, we will derive exactly the same expression (\ref{eq:energy mercury})
for the energy of Mercury's orbit via the present Machian approach.
The celestial system is modeled as a 3-body problem between Mercury,
the sun and the rest of the universe, which we represent by the hollow
sphere $O$ (accounting for the background potential $\varphi_{o}$).
According to (\ref{eq:E}), the total energy $E_{o}$ of the 3-body
system ($m$,$M$,$O$) is
\begin{equation}
E_{o}=T_{mO}+T_{mM}+V_{mO}+V_{mM}={\scriptstyle \frac{1}{2}}(m+\mu)\dot{r}^{2}+{\scriptstyle \frac{1}{2}}mr^{2}\dot{\phi}^{2}+(m+\mu)\varphi_{o}\label{eq:E orbit merc}
\end{equation}
where
\begin{equation}
\mu=\mu_{m{\scriptscriptstyle M}}=\frac{{\scriptstyle -}GmM}{\varphi_{o}r}.\label{eq:mu}
\end{equation}
Note that (\ref{eq:E orbit merc}) matches Schrödinger's result in
\cite{Schroedinger (1926)}. Via a Lagrangian approach, we obtain
from (\ref{eq:E orbit merc}) the momentum of the orbit (per unit
mass) as a constant of motion
\begin{equation}
h=r^{2}\dot{\phi}.\label{eq:h orbit}
\end{equation}
Substituting this in (\ref{eq:E orbit merc}), and multiplying both
sides by $m/(m+\mu)$ gives
\begin{equation}
{\scriptstyle \frac{1}{2}}m\dot{r}^{2}+{\scriptstyle \frac{1}{2}}\frac{m}{m+\mu}\frac{mh^{2}}{r^{2}}+m\varphi_{o}=\frac{m}{m+\mu}E_{o}.\label{eq:E orbit merc-1}
\end{equation}
Introducing the auxiliary constant $\epsilon_{o}$, the constant total
energy $E_{o}$ may be defined as
\begin{equation}
E_{o}=m\varphi_{o}+\epsilon_{o}.\label{eq:Eo}
\end{equation}
Note that for sub-relativistic speeds $E_{o}\approx m\varphi_{o}={\scriptstyle -}{\scriptstyle \frac{1}{2}}mc^{2}$
or
\begin{equation}
\epsilon_{o}\ll m\varphi_{o}.\label{eq:eps}
\end{equation}
Since $\mu\ll m$, we may replace in (\ref{eq:E orbit merc-1}) $m/(m+\mu)$
by $(m-\mu)/m$. So ,using (\ref{eq:Eo}), we get
\begin{equation}
{\scriptstyle \frac{1}{2}}m\dot{r}^{2}+{\scriptstyle \frac{1}{2}}(m-\mu)\frac{h^{2}}{r^{2}}+m\varphi_{o}=\frac{m{\scriptstyle -}\mu}{m}E_{o}=\epsilon_{o}+m\varphi_{o}-\frac{\mu}{m}\epsilon_{o}-\mu\varphi_{o}.\label{eq:E orbit merc-1-1}
\end{equation}
Then, moving the constant energy terms to the right side, ignoring
$\frac{\mu}{m}\epsilon_{o}$ (since (\ref{eq:eps})) and substituting
(\ref{eq:mu}) yields
\begin{equation}
{\scriptstyle \frac{1}{2}}m\dot{r}^{2}+{\scriptstyle \frac{1}{2}}m\frac{h^{2}}{r^{2}}-\frac{GmM}{c^{2}}\frac{h^{2}}{r^{3}}-\frac{GmM}{r}=\epsilon_{o}\label{eq:H orbit}
\end{equation}
by which we have retrieved the relativistic energy equation of Mercury's
orbit (\ref{eq:energy mercury}), from which of course the same expression
(\ref{eq:GR Anomalous prec}) follows for the anomalous precession.
This result supports the view that inertia is anisotropic, in agreement
with Berkeley. In addition, it is pointed out in Appendix A that an
isotropic model can not explain the anomalous precession.

\subsection{Frame-dragging\label{sub:Frame-dragging}}

Perhaps the most convincing test of the Machian approach is its simplicity
in the derivation of frame-dragging effects. We consider the simplest
form, linear frame-dragging, followed by rotational (Lense-Thirring)
frame-dragging.

\subsubsection*{Linear frame-dragging}

We study the behavior of a small test particle of mass $m$ near a
larger mass $M$ against the background potential $\varphi_{o}$ of
universe $O$. If no (net) forces are exerted on the particle, then
it will remain still in it's local frame, so the particle's position
marks the position of the frame. If we let $M$ be a long tube, and
place the particle $m$ halfway inside the tube, exactly on the tube's
axis, then the particle will be weightless, i.e., the net force of
gravity on the particle is zero. Let $x_{m}(0)$ be the initial position
of $m$ and let $v_{m}(0)=0$ and $v_{{\scriptscriptstyle M}}(0)=0$
be the initial velocities of $m$ and $M$, respectively. At some
point, we make the tube move in the axial direction with a uniform
velocity $v_{{\scriptscriptstyle M}}\neq0$. In the Newtonian context,
this would not affect the position of $m$, so $v_{m}$ remains zero.
Formally: the Newtonian total kinetic energy is $T_{Newton}={\scriptstyle \frac{1}{2}}Mv_{{\scriptscriptstyle M}}^{2}+{\scriptstyle \frac{1}{2}}mv_{m}^{2}$,
and the linear momentum of $m$: $p_{Newton}=\partial T_{Newton}/\partial v_{m}=mv_{m}$,
which tells that if $v{}_{m}$ is constant and zero, then $p_{Newton}=0$
and the net force on $m$, $F_{m}=\dot{p}_{m}=0$ will also remain
zero.\medskip{}

\noindent In the Machian approach, however, we have the following
expression for the total kinetic energy
\begin{equation}
T_{Mach}={\scriptstyle \frac{1}{2}}Mv_{{\scriptscriptstyle M}}^{2}+{\scriptstyle \frac{1}{2}}mv_{m}^{2}+{\scriptstyle \frac{1}{2}}\mu(v_{m}{\scriptstyle -}v_{{\scriptscriptstyle M}})^{2}\label{eq:Tmach}
\end{equation}
where $\mu=$$\mu_{mM}=m\varphi_{{\scriptscriptstyle M}}(x_{m})/\varphi_{o}$.
The last term in (\ref{eq:Tmach}) represents the kinetic energy of
the particle relative to the tube, which term is absent in the classical
kinetic energy $T_{Newton}$. It is this last term that links the
particle and the tube, energy-wise. Also in the Machian approach,
$m$ is not acted upon, so we still have zero total momentum associated
with the particle's axial velocity $v_{m}$
\begin{equation}
p_{Mach}=\frac{\partial T_{Mach}}{\partial v_{m}}=mv_{m}+\mu(v_{m}{\scriptstyle -}v_{{\scriptscriptstyle M}})=0\label{eq:pm}
\end{equation}
from which we obtain the particle's velocity as a function of the
tube's velocity
\begin{equation}
v_{m}=\frac{\mu v_{{\scriptscriptstyle M}}}{m+\mu}=\frac{\varphi_{{\scriptscriptstyle M}}(x_{m})}{\varphi_{o}+\varphi_{{\scriptscriptstyle M}}(x_{m})}\, v_{M}.\label{eq:linear drag}
\end{equation}
In other words: \textit{without exerting any force} on it ($p_{Mach}=0$
at any point in time), the particle seems to start moving with a fraction
of the tube's velocity. This is called \textit{frame dragging}. However,
as pointed out, there actually is no dragging force involved as long
as $p_{Mach}=0$. This notion is in full agreement with the following
interpretation: the acceleration of the tube from standstill to some
finite velocity $v_{{\scriptscriptstyle M}}$ causes the universe
$O$, due to the reaction force, to accelerate from standstill to
a velocity ${\scriptscriptstyle {\scriptstyle -}}v_{m}$ in the opposite
direction of $v_{{\scriptscriptstyle M}}$. The tube seems to drag
the frame, but from the Machian point of view, the frame and the particle
remain at rest and the tube's velocity is actually $(v_{{\scriptscriptstyle M}}{\scriptstyle -}v_{m})$.
Hence, the particle can be considered to not move at all, in agreement
with $p_{Mach}=0$. Furthermore, note that (\ref{eq:pm}) shows that
the total linear momentum of $m$ is the sum of the particle's partial
linear momenta
\begin{equation}
p_{Mach}=p_{mO}+p_{mM}=mv_{m}+\mu(v_{m}{\scriptstyle -}v_{{\scriptscriptstyle M}})=0.\label{eq:p vm}
\end{equation}
The other interesting observation about (\ref{eq:linear drag}) is
that, for arbitrary background potential $\varphi_{b}$, the 'dragging
fraction' at position $x$ is equal to the ratio $\varphi_{{\scriptscriptstyle M}}(x)/(\varphi_{b}+\varphi_{{\scriptscriptstyle M}}(x))$.
This means that in the limit of $\varphi_{b}{\scriptstyle \rightarrow}0$
(when we get into empty space) the fraction approaches unity, hence,
$v_{m}{\scriptstyle \rightarrow}v_{{\scriptscriptstyle M}}$. So the
particle moves completely along with the tube, which makes perfectly
sense from a Machian point of view, because any common velocity of
$m$ and $M$ in empty space does not represent kinetic energy, is
artificial and is equivalent to no motion at all. If, however, $\varphi{}_{b}=\varphi_{o}$,
as near earth, then the ratio is extremely small, i.e., hard to verify
experimentally.

\subsubsection*{Lense-Thirring frame dragging}

The Lense-Thirring effect (\cite{Lense Thirring 1918} and \cite{Iorio (2011)}
for a recent review) concerns frame-dragging near a rotating sphere,
say planet earth. Let $M$ and $\omega_{{\scriptscriptstyle M}}$
denote the mass and spin angular velocity of the earth. An infinitesimal
small particle of mass $m$ in the vicinity of $M$ will be \textquotedbl{}dragged\textquotedbl{}
by $M$ in a circular orbit around $M$ with an angular velocity $\dot{\phi}$,
which is a fraction of the angular velocity of the sphere. To prevent
the particle from falling to $M$, we assume the particle is constrained
by a circular polar orbit of radius $r_{m}$ in the $r,\theta$ plane.
Thus, the dragging of $m$ results in a (slow) precession $\dot{\phi}$
of the plane of the orbit. We position the origin of the polar coordinates
onto the CM of the two spheres. Since $m$ is infinitesimal small,
the CM is at the center of $M$, so $r_{m}=r$ and $r_{{\scriptscriptstyle M}}=0$.
The derivation via GR, involving the Kerr-metric, yields the well-known
expression $\dot{\phi}_{{\scriptscriptstyle GR}}$ for the Lense-Thirring
precession of particle $m$
\begin{equation}
\dot{\phi}_{{\scriptscriptstyle GR}}=\frac{r_{s}\lambda rc}{(r^{2}+\lambda^{2}\cos^{2}\theta)(r^{2}+\lambda^{2})+r_{s}\lambda^{2}r\sin^{2}\theta},\label{eq:omega LT}
\end{equation}
where $r_{s}=2GM/c^{2}$ is the Schwarzschild radius, $\lambda=J/Mc$
a distance scale and $J={\scriptscriptstyle \frac{2}{5}}MR_{{\scriptscriptstyle M}}^{2}\omega_{{\scriptscriptstyle M}}$
the angular momentum of $M$. In case of planet earth, $\lambda\ll r$
and $r{}_{s}\ll r$ for any $r\geq R_{{\scriptscriptstyle M}}$. Thus,
(\ref{eq:omega LT}) may be simplified to
\begin{equation}
\dot{\phi}_{{\scriptscriptstyle GR}}=\frac{r_{s}\lambda c}{r^{3}}=\frac{2GM{\scriptscriptstyle \frac{2}{5}}R_{{\scriptscriptstyle M}}^{2}}{c^{2}r^{3}}\,\omega_{{\scriptscriptstyle M}}\label{eq:omega GR}
\end{equation}
which can be rewritten as
\begin{equation}
\dot{\phi}_{{\scriptscriptstyle GR}}=\frac{\varphi_{{\scriptscriptstyle M}}(r){\scriptscriptstyle \frac{2}{5}}R_{{\scriptscriptstyle M}}^{2}}{\varphi_{o}r^{2}}\,\omega_{{\scriptscriptstyle M}}.\label{eq:omega GR-1}
\end{equation}
 Now, we will start from the other end and derive the Machian equivalent
of this GR based expression (\ref{eq:omega GR-1}). The Machian Lagrangian
for the system reads 
\begin{equation}
\mathcal{L}_{Mach}=T_{Mach}-V_{Mach}\label{eq:L Lense-Thirring}
\end{equation}
\[
T_{Mach}=T_{mM}+T_{mO}+T_{MO}=(T_{\dot{r}_{mM}}+T_{\omega_{mM}})+(T_{v_{mO}}+T_{\omega_{mO}})+(T_{v_{MO}}+T_{\omega_{MO}})
\]
\[
V_{Mach}=V_{mM}+V_{mO}+V_{MO}
\]
In analogy with linear frame dragging, we only have to derive the
expression for the total angular momentum of particle $m$ associated
with $\dot{\phi}$, set this to zero and solve for $\dot{\phi}$:
\begin{equation}
J_{\dot{\phi}}=\frac{\partial\mathcal{L}_{Mach}}{\partial\dot{\phi}_{GR}}=0.\label{eq:p omega}
\end{equation}
For this purpose, we skip all contributions to $\mathcal{L}{}_{Mach}$
which are independent of $\dot{\phi}$ and keep the energy terms associated
with $\dot{\phi}$, i.e., the Machian term $T_{\omega_{mM}}$ of the
spin energies between $m$ and $M$ (between square brackets) and
the Newtonian terms $T_{v_{mO}}$ and $T_{v_{MO}}$ of the precessing
motion of $m$ respectively $M$ relative to the universe:
\begin{equation}
\mathcal{L}_{\dot{\phi}}=\left[{\scriptstyle \frac{1}{5}}\mu R_{m}^{2}(\omega_{m}{\scriptstyle -}\dot{\phi})^{2}+{\scriptstyle \frac{1}{5}}\mu R_{{\scriptscriptstyle M}}^{2}(\omega_{{\scriptscriptstyle M}}{\scriptstyle -}\dot{\phi})^{2}\right]\:+\:{\scriptstyle \frac{1}{2}}mr_{m}^{2}\dot{\phi}{}^{2}\:+\:{\scriptstyle \frac{1}{2}}Mr_{{\scriptscriptstyle M}}^{2}\dot{\phi}{}^{2}\label{eq:L omega}
\end{equation}
where $\mu=-GmM/\varphi_{o}r$. The first part of the Machian term
in (\ref{eq:L omega}) is the spin energy of the particle $m$ relative
to $M$. This part is zero, because the particle's radius $R_{m}$
is infinitesimal small. The second part of the Machian term is the
spin energy of $M$ relative to the particle, which is (relatively)
significant. The third term represents the Newtonian energy of the
particle's precession relative to the universe, which energy is also
significant. The last term represents the Newtonian energy of the
precession of $M$ relative to the universe. This term can be omitted,
since $r_{{\scriptscriptstyle M}}=0$. This leaves
\begin{equation}
\mathcal{L}_{\dot{\phi}}={\scriptstyle \frac{1}{5}}\mu R_{{\scriptscriptstyle M}}^{2}(\omega_{{\scriptscriptstyle M}}{\scriptstyle -}\dot{\phi})^{2}+{\scriptstyle \frac{1}{2}}mr^{2}\dot{\phi}^{2}\label{eq:L omega-1}
\end{equation}
where we replaced $r_{m}$ by $r$. So the momentum of particle $m$
associated with angular velocity $\dot{\phi}$ is
\begin{equation}
J_{\dot{\phi}}=\frac{\partial\mathcal{L}_{Mach}}{\partial\dot{\phi}}=\frac{\partial\mathcal{L}_{\dot{\phi}}}{\partial\dot{\phi}}={\scriptstyle -}{\scriptstyle \frac{2}{5}}\mu R_{{\scriptscriptstyle M}}^{2}(\omega_{{\scriptscriptstyle M}}{\scriptstyle -}\dot{\phi})+mr^{2}\dot{\phi}=0\label{eq:p omega-1}
\end{equation}
from which we obtain (labeling $\dot{\phi}=\dot{\phi}_{{\scriptscriptstyle Mach}}$)
\begin{equation}
\dot{\phi}_{{\scriptscriptstyle Mach}}=\frac{{\scriptstyle \frac{2}{5}}\mu R_{{\scriptscriptstyle M}}^{2}\omega_{{\scriptscriptstyle M}}}{mr^{2}+{\scriptstyle \frac{2}{5}}\mu R_{{\scriptscriptstyle M}}^{2}}=\frac{\varphi_{{\scriptscriptstyle M}}(r)\,{\scriptstyle \frac{2}{5}}R_{{\scriptscriptstyle M}}^{2}}{\varphi_{o}r^{2}+\varphi_{{\scriptscriptstyle M}}(r){\scriptstyle \frac{2}{5}}R_{{\scriptscriptstyle M}}^{2}}\,\omega_{{\scriptscriptstyle M}}\approx\frac{\varphi_{{\scriptscriptstyle M}}(r)\,{\scriptstyle \frac{2}{5}}R_{{\scriptscriptstyle M}}^{2}}{\varphi_{o}r^{2}}\,\omega_{{\scriptscriptstyle M}}=\dot{\phi}_{{\scriptscriptstyle GR}},\label{eq:omega mach}
\end{equation}
i.e. the Machian expression matches the GR expression (\ref{eq:omega GR-1}):
the difference is negligible, as $\varphi_{{\scriptscriptstyle M}}(r)\ll\varphi_{o}$
and $R{}_{{\scriptscriptstyle M}}<r$. Note that the expression (\ref{eq:p omega-1})
for $J_{\dot{\phi}}$ allows for a Machian interpretation analog to
linear frame-dragging: from the point of view of the particle, earth
is rotating with an angular velocity $\omega_{{\scriptscriptstyle M}}{\scriptstyle -}\dot{\phi}$,
which is compensated by a backward rotation ${\scriptstyle -}\dot{\phi}$
of the rest of the universe, in such a way that the total momentum
$J_{\dot{\phi}}$ remains zero. This interpretation agrees with \cite{Iorio (2011)},
where Iorio et al. point out that the gravitomagnetic field has no
physical existence, yet causes Lense-Thirring precession as \textquotedbl{}a
pure coordinate artifact\textquotedbl{}.

\section{The relational model as an isotropic transform of the anisotropic
Machian model\label{sec:The-relation-model}}

\subsection{Remote observation by the isotropic relational metric}

\noindent The above examples demonstrate the validity of the anisotropic
Machian model in predicting relativistic trajectories of dynamical
systems. Notably, this did not involve the spacetime concept. The
isotropic model, on the other hand, explains relativistic remote observation
phenomena such as time dilation, as pointed out in section \ref{sub:Isotropic-model}.
We assume the observer is essentially a local reference frame in spacetime
and its presence does not influence the system under study. In curved
spacetime, the change of the observer's position affects his measurements
of the system. In the relational view, units vary along with local
potential, i.e. the relational metric is a function of potential.
We will derive such a relational metric from the Schwarzschild metric
and the corresponding Machian energy equation for a test particle
$m$, orbiting a large sphere of mass $M$ against a background potential
$\varphi_{o}$. This result will then be generalized for arbitrary
spacetimes.\medskip{}

\noindent The Schwarzschild metric in spherical coordinates ($r$,$\phi$,$\theta$)
reads
\begin{equation}
d\lambda^{2}=c^{2}d\tau^{2}=\alpha_{s}c^{2}dt^{2}-\frac{dr^{2}}{\alpha_{s}}-r^{2}d\theta^{2}-r^{2}\sin\theta d\phi^{2}.\label{eq:Schwarzschild metric}
\end{equation}
The Schwarzschild dilation parameter is defined $\alpha_{s}=1{\scriptstyle -}r_{s}/r$,
where $r_{s}=2GM/c^{2}$ is the Schwarzschild radius. We again assume
the system has a background potential $\varphi_{o}$ (see the note
on the Schwarzschild vacuum in section \ref{sub:Isotropic-model}).
This is not an arbitrary choice, since, according to (\ref{eq:phi o}),
$\varphi_{o}={\scriptstyle -}{\scriptstyle \frac{1}{2}}c^{2}$. The
potential represents the speed of light in the Machian context. Or,
conversely, the speed of light represents the background potential
$\varphi_{o}$ in the Schwarzschild metric. This relationship connects
both metrics. Choosing axes such that $\theta=\pi/2$ in the plane
of the orbit simplifies the metric to
\begin{equation}
c^{2}d\tau^{2}=\alpha_{s}c^{2}dt^{2}-\frac{dr^{2}}{\alpha_{s}}-r^{2}d\phi^{2}.\label{eq:S-metric planar-1}
\end{equation}
Taking on both sides the derivative with respect to coordinate time
($\frac{d}{dt}$ denoted by the dot) yields the differential equation
for the orbit in coordinate time
\begin{equation}
c^{2}\dot{\tau}^{2}=\alpha_{s}\, c^{2}-\frac{\dot{r}^{2}}{\alpha_{s}}-r^{2}\dot{\phi}^{2}\label{eq:coordinate time metric}
\end{equation}
Next, we isolate the constant energy (per unit mass) $c^{2}$ from
the first term on the right
\begin{equation}
c^{2}=\frac{c^{2}\dot{\tau}^{2}}{\alpha_{s}}+\frac{\dot{r}^{2}}{\alpha_{s}^{2}}+\frac{r^{2}\dot{\phi}^{2}}{\alpha_{s}}\label{eq:coordinate time metric-1}
\end{equation}
from which equation we identify the constant of motion
\begin{equation}
\frac{\dot{\tau}}{\alpha_{s}}=k=const,\label{eq:tau tc}
\end{equation}
by which $\dot{\tau}$ can be eliminated from (\ref{eq:coordinate time metric-1}),
yielding the following equation in coordinate time
\begin{equation}
c^{2}=\alpha_{s}c^{2}k^{2}+\frac{\dot{r}^{2}}{\alpha_{s}^{2}}+\frac{r^{2}\dot{\phi}^{2}}{\alpha_{s}},\label{eq:coordinate time geo}
\end{equation}
which we may rearrange into
\begin{equation}
\frac{{\scriptstyle -}c^{2}}{\alpha_{s}}+\frac{\dot{r}^{2}}{\alpha_{s}^{3}}+\frac{r^{2}\dot{\phi}^{2}}{\alpha_{s}^{2}}=-c^{2}k^{2}.\label{eq:coordinate time geo-1}
\end{equation}
Multiplying both sides by ${\scriptstyle \frac{1}{2}}m$ and using
$\varphi_{o}=-{\scriptstyle \frac{1}{2}}c^{2}$ yields the \textquotedbl{}Schwarzschild
energy equation\textquotedbl{} for the orbit
\begin{equation}
\frac{m\varphi_{o}}{\alpha_{s}}+\frac{{\scriptstyle \frac{1}{2}}m\dot{r}^{2}}{\alpha_{s}^{3}}+\frac{{\scriptstyle \frac{1}{2}}mr^{2}\dot{\phi}^{2}}{\alpha_{s}^{2}}={\scriptstyle -}{\scriptstyle \frac{1}{2}}mc^{2}k^{2}=E_{o}.\label{eq:Schwarz energy eq}
\end{equation}
Like in section \ref{sub:Isotropic-model}, we consider the Schwarzschild
dilation factor $\alpha_{{\scriptscriptstyle S}}=1-\frac{r_{s}}{r}$
as a special case of the more general relational dilation factor $\alpha_{{\scriptscriptstyle R}}$
between two arbitrary positions $x_{A}$ and $x_{B}$
\begin{equation}
\alpha_{{\scriptscriptstyle R}}(x_{A},x_{B})=\frac{\varphi(x_{A})}{\varphi(x_{B})}\label{eq:alpha M}
\end{equation}
which, in the present case of the Schwarzschild spacetime, i.e. for
a particle $m$ at position $r$ and an observer at infinity (where
we assume a background potential $\varphi_{o}$), takes on the particular
value $\hat{\alpha}_{{\scriptscriptstyle R}}$
\begin{equation}
\hat{\alpha}_{{\scriptscriptstyle R}}=\alpha_{{\scriptscriptstyle R}}(\infty,r)=\frac{\varphi_{o}}{\varphi_{o}+\varphi_{{\scriptscriptstyle M}}(r)}=\frac{1}{1+\frac{r_{s}}{r}}\approx1{\scriptstyle -}\frac{r_{s}}{r}=\alpha_{s}\label{eq:alpha M Schwarzschild}
\end{equation}
Replacing in (\ref{eq:Schwarz energy eq}), like in section \ref{sub:Isotropic-model},
$\alpha_{s}$ by $\hat{\alpha}_{{\scriptscriptstyle R}}$ and recalling
relations $\mu=\mu_{mM}={\scriptstyle -}m\varphi_{{\scriptscriptstyle M}}(r)/\varphi_{o}$,
$m=\mu_{om}$ and $m/\hat{\alpha}_{{\scriptscriptstyle R}}=m+\mu$
yields the following \textquotedbl{}relational energy equation\textquotedbl{}
in \textquotedbl{}coordinate\textquotedbl{} time $t$
\begin{equation}
(m+\mu)\varphi_{o}+\frac{{\scriptstyle \frac{1}{2}}(m+\mu)\dot{r}^{2}}{\hat{\alpha}_{{\scriptstyle {\scriptscriptstyle R}}}^{2}}+\frac{{\scriptstyle \frac{1}{2}}mr^{2}\dot{\phi}^{2}}{\hat{\alpha}_{{\scriptstyle {\scriptscriptstyle R}}}^{2}}={\scriptstyle -}{\scriptstyle \frac{1}{2}}mc^{2}k^{2}=E_{o}\label{eq:relational energy eq}
\end{equation}
This is the Machian energy equation of the orbit (\ref{eq:E}), except
for the common denominator $\hat{\alpha}_{{\scriptstyle {\scriptscriptstyle R}}}^{2}$
of the two kinetic energy terms, which points at an \textit{isotropic
coordinate transform} between the observer's coordinates at infinity
($r$,$s_{\phi}$,$t$) and the local coordinates ($\rho,\sigma_{\phi},\tau$)
of the test particle (where we let $s_{\phi}=r\phi$ and $\sigma_{\phi}=\rho\phi$
represent the tangential coordinate in the two frames). Eq. (\ref{eq:relational energy eq})
in ($r$,$s_{\phi}$,$t$) coordinates reads
\begin{equation}
(m+\mu)\varphi_{o}+\frac{{\scriptstyle \frac{1}{2}}(m+\mu)\dot{r}^{2}}{\hat{\alpha}_{{\scriptstyle {\scriptscriptstyle R}}}^{2}}+\frac{{\scriptstyle \frac{1}{2}}m\dot{s}_{\phi}^{2}}{\hat{\alpha}_{{\scriptstyle {\scriptscriptstyle R}}}^{2}}=E_{o}\label{eq:Orbit (r,s,t)}
\end{equation}
If we rewrite the Machian energy equation (\ref{eq:E}) in local coordinates
($\rho,\sigma_{\phi},\tau$) (the derivative $\frac{d}{d\tau}$ is
denoted by the circle ${\scriptscriptstyle ^{\circ}}$):
\begin{equation}
(m+\mu)\varphi_{o}+{\scriptstyle \frac{1}{2}}(m+\mu)\overset{{\scriptscriptstyle \circ}}{\rho}\phantom{}^{2}+{\scriptstyle \frac{1}{2}}m\overset{{\scriptscriptstyle \circ}}{\sigma}\phantom{}_{\phi}^{2}=E_{o},\label{eq:Mach local}
\end{equation}
then equating (\ref{eq:Orbit (r,s,t)}) and (\ref{eq:Mach local})
yields for the radial direction
\begin{equation}
\overset{{\scriptscriptstyle \circ}}{\rho}\phantom{}^{2}=\frac{\dot{r}^{2}}{\hat{\alpha}_{{\scriptstyle {\scriptscriptstyle R}}}^{2}}\label{eq:transf A}
\end{equation}
and identically for the tangential direction
\begin{equation}
\overset{{\scriptscriptstyle \circ}}{\sigma}\phantom{}_{\phi}^{2}=\frac{\dot{s}_{\phi}^{2}}{\hat{\alpha}_{{\scriptstyle {\scriptscriptstyle R}}}^{2}}\label{eq:transf B}
\end{equation}
Thus, assuming the validity of both the Schwarzschild metric and the
corresponding Machian energy equation, remote observation causes a
coordinate transform of velocity. From this velocity transform, we
infer the underlying relational spacetime metric for the Schwarzschild
case as follows: recalling (\ref{eq:matching}), gravitational time
dilation of local time v.s. coordinate time is expressed as
\begin{equation}
d\tau^{2}=\hat{\alpha}_{{\scriptscriptstyle R}}dt^{2}.\label{eq:time dilation}
\end{equation}
Combining (\ref{eq:time dilation}) with (\ref{eq:transf A}) and
(\ref{eq:transf B}), respectively, yields $d\rho^{2}=\frac{1}{\hat{\alpha}_{{\scriptstyle {\scriptscriptstyle R}}}}dr^{2}$
and, identically, $d\sigma_{\phi}^{2}=\frac{1}{\hat{\alpha}_{{\scriptstyle {\scriptscriptstyle R}}}}ds_{\phi}^{2}$.
Because of this spatial isotropy, we need not discern between the
radial and tangential directions and consider only a single spatial
metric, which applies to a spatial coordinate $s$ in any direction,
so to both $dr$ and $ds_{\phi}=rd\phi$ according to
\begin{equation}
d\sigma^{2}=\frac{1}{\hat{\alpha}_{{\scriptstyle {\scriptscriptstyle R}}}}ds^{2}\label{eq:spatial metric}
\end{equation}
where $\sigma$ is the local coordinate and $s$ the coordinate at
infinity. Hence, the relational equivalent of the Schwarzschild metric
presumes identical length contraction in both radial and tangential
direction, i.e. isotropic contraction. Where the Schwarzschild metric
has anisotropic length contraction, this anisotropic feature is represented
in the Machian energy equation (\ref{eq:Mach local}), and consequently
in the relational energy equation (\ref{eq:relational energy eq}),
by the additional inertia into the radial direction:
\begin{equation}
m+\mu=\frac{m}{\hat{\alpha}_{{\scriptstyle {\scriptscriptstyle R}}}}\label{eq:inertia}
\end{equation}
Hence, by mapping the anisotropic Machian model onto the anisotropic
Schwarzschild metric we distilled an isotropic relational metric for
the Schwarzschild case.

\subsection{General form of the relational metric\label{sub:General-form-of}}

The specific relational metric (\ref{eq:time dilation}),(\ref{eq:spatial metric})
for the Schwarzschild case expresses the coordinate transform between
local coordinates of the particle $m$ and the coordinates of the
observer at infinity. Actually, considering definition $\hat{\alpha}_{{\scriptscriptstyle R}}=\varphi_{o}/\varphi(r)$,
this specific metric expresses the ratio of local units at two specific
potentials. This naturally suggests the generalization of the metric
to arbitrary potentials of arbitrary spacetimes by taking, instead
of $\hat{\alpha}_{{\scriptscriptstyle R}}$, the general form $\alpha_{{\scriptscriptstyle R}}$
(\ref{eq:alpha M}) of the relational dilation factor between two
arbitrary potentials. Hence, the coordinates ($\sigma_{A},\tau_{A}$)
and ($\sigma_{B},\tau_{B}$) at positions $x_{A}$ and $x_{B}$ relate
according to
\begin{equation}
d\tau_{B}^{2}=\alpha_{{\scriptscriptstyle R}}d\tau_{A}^{2}\label{eq:dt dilation}
\end{equation}
\begin{equation}
d\sigma_{B}^{2}=\frac{1}{\alpha_{{\scriptscriptstyle R}}}d\sigma_{A}^{2}\label{eq:ds contr}
\end{equation}
or, alternatively
\begin{equation}
\varphi(x_{B})\, d\tau_{B}^{2}=\varphi(x_{A})\, d\tau_{A}^{2}\label{eq:dt dilation-1}
\end{equation}
\begin{equation}
\varphi(x_{A})\, d\sigma_{B}^{2}=\varphi(x_{B})\, d\sigma_{A}^{2}\label{eq:ds contr-1}
\end{equation}
Eq. (\ref{eq:dt dilation-1}) and (\ref{eq:ds contr-1}) present a
general relational metric, satisfying the relational principle: suppose,
from earth, at position $x_{A}$, we observe an infinitesimal small
test particle entering empty space, i.e. getting farther away from
earth and all other masses. The local potential $\varphi(x_{B})$
at the position of the particle will gradually vanish and, as a result,
the local coordinates ($\sigma_{B},\tau_{B}$) of the particle dilute,
meaning unbound increase of the local clock rate and of the unit length,
while moving towards vacuum infinity.

\subsection{The relational model}

The \textit{relational model} consists of the Machian model subject
to the coordinate transform, defined by the relation metric (\ref{eq:dt dilation-1}),(\ref{eq:ds contr-1}).
The relational model fits the relational principle (including Mach's
principle) and appears consistent with GR, provided the background
potential is equal to $\varphi_{o}$. Under that condition we retrieve
the Schwarzschild dilation factor ($\alpha_{{\scriptscriptstyle R}}=\hat{\alpha}_{{\scriptscriptstyle R}}\approx\alpha_{{\scriptscriptstyle S}}$).
Since the relational model matches GR in the Schwarzschild case, it
matches major GR results: gravitational time dilation and redshift,
deflection of a photon by the sun, anomalous perihelion precession.
The relational model, however, applies to general spacetimes, i.e.
is not limited to the Schwarzschild spacetime. One example is Lense-Thirring
frame-dragging (section \ref{sub:Frame-dragging}), which regards
the Kerr spacetime. The real proof of concept is likely to be found
on the galactic or cosmic scale.

\subsection{Cosmological implications\label{sub:Cosmological-implications}}

When in the Schwarzschild case the background potential gets significantly
below $\varphi_{o}$, the corresponding relational model starts to
deviate significantly from the Schwarzschild metric, since at lower
potential the relational metric will expand spacetime, while the Schwarzschild
metric will remain unchanged. This may shed a light on some questions
of remote observation on the galactic and cosmic scale, such as the
galaxy rotation curve problem and the question of the accelerated
expansion of the universe.\medskip{}

\noindent \textit{The galaxy rotation problem} - This regards the
flat rotation velocity curve of stars on the outskirts of a galaxy.
Their speed does not decrease at increasing distance from the galactic
center as one would expect by Newton's laws, i.e. stars seem to move
too fast. Considering the relational metric, the lower potential away
from the central bulge has several implications: the inertia of a
star relative to the central bulge mass is less, the local meter near
the star is longer and the local time runs faster. All these factors
cause a higher value of the \textit{observed} velocity relative to
the expected Newtonian velocity. \medskip{}

\noindent \textit{Accelerated expansion of the universe }- Objects
receding from us into areas of weaker potential lose kinetic energy,
however gain speed (as observed from earth) due to decreasing inertia
and the expanded spacetime at weaker potentials. Consider an infinitesimal
small test particle $m$ entering empty space, moving away from the
total universal mass, represented by a sphere $M$ of radius $R_{{\scriptscriptstyle M}}$.
We assume the mass density inside of $M$ is spherically symmetric.
The particle is at a distance $r>R_{{\scriptscriptstyle M}}$ from
the center of $M$, where $r$ is measured in the local length unit
of the particle. The particle is observed from earth, so the observer
is situated inside of $M$ at potential $\varphi_{o}$ and the relational
dilation factor for the observer relative to the position of the particle
is
\begin{equation}
\alpha_{{\scriptscriptstyle R}}=\frac{{\scriptstyle -}\varphi_{o}r}{GM}
\end{equation}
which grows unbound for $r{\scriptstyle \rightarrow}\infty$. Recalling
(\ref{eq:T and V 2-body empty}), the total energy of the two body
system $(M,m)$ is
\begin{equation}
E={\scriptscriptstyle \frac{1}{2}}\mu\dot{r}^{2}+\mu\varphi_{o}={\scriptscriptstyle \frac{1}{2}}\mu\dot{r}^{2}-{\scriptscriptstyle \frac{1}{2}}\mu c^{2}\label{eq:E two body}
\end{equation}
where $\mu\varphi_{o}\leq E\leq0$ and $\mu=-GmM/\varphi_{o}r$, from
which we obtain the (squared) velocity of the particle as a function
of distance $r$
\begin{equation}
\dot{r}^{2}=\frac{2E}{\mu}-2\varphi_{o}=c^{2}\left(\frac{E}{GmM}\cdot r+1\right).\label{eq:(dr/dt)2}
\end{equation}
For $r>R_{{\scriptscriptstyle M}}$ the particle will decelerate (since
$E$ is negative) and will reach a maximum distance $r_{max}$ at
the point where $\dot{r}=0$, i.e.
\begin{equation}
r_{max}=\frac{{\scriptstyle -}GmM}{E},
\end{equation}
and will eventually fall back to $M$. By the metric (\ref{eq:dt dilation}),(\ref{eq:ds contr})
the speed of the particle, but as observed from earth, is equal to
\begin{equation}
\dot{r}_{obs}^{2}=\alpha_{{\scriptscriptstyle R}}^{2}\dot{r}^{2}=\alpha_{{\scriptscriptstyle R}}^{2}c^{2}\left(\frac{E}{GmM}\cdot r+1\right)=c^{2}\left(\frac{\varphi_{o}}{GM}\right)^{2}r^{2}\left(\frac{E}{GmM}\cdot r+1\right).\label{eq:(drobs/dt)2}
\end{equation}
So when $m$ enters empty space, $\alpha_{{\scriptscriptstyle R}}$
grows and the particle will at first appear to accelerate, as seen
from earth. A maximum observed velocity will be reached where
\begin{equation}
\frac{d(\dot{r}_{obs}^{2})}{dr}=c^{2}\left(\frac{\varphi_{o}}{GM}\right)^{2}\left(\frac{E}{GmM}\cdot3r^{2}+2r\right)=0,
\end{equation}
which is at a distance
\begin{equation}
\hat{r}={\scriptstyle -}{\scriptscriptstyle \frac{2}{3}}\frac{GmM}{E}={\scriptscriptstyle \frac{2}{3}}r_{max}.\label{eq:r max obs velocity}
\end{equation}
It is only beyond this point, at 2/3 of its total outward journey,
that also in the eye of the observer the particle will decelerate,
until standstill at $r_{max}$ and fall back. Apparently, and to our
comfort, the universe is not yet at this stage and is still in accelerating
phase.

\section{Concluding notes}

In the tradition of Descartes, Newton, Berkeley, Mach, Einstein and
Schrödinger, we discussed a number of concepts of relational physics
and demonstrated the feasibility of a relational theory of relativity.
To some extend, one can view this as a reinterpretation of GR by a
rotation of concepts. However, the relational concepts have different
roots and a merit of their own. In summary: 
\begin{lyxlist}{00.00.0000}
\item [{{a)}}] The principle of Berkeley: only radial motion between
two (point) masses in otherwise empty space represents inertia and
kinetic energy between these masses. This is essentially the basis
for the anisotropic Machian model. Moreover, it makes the emergence
of space and time from the interaction of masses conceivable. 
\item [{{b)}}] The notion that inertia (Mach) and kinetic energy (Schrödinger)
are mutual properties between pairs of objects, like potential energy,
leading to Machian definitions of inertia and (frame independent)
kinetic energy, which necessarily reflect the definition of potential
energy, in order to provide conservation of energy under arbitrary
choice of frame. This provides also for the definition of the Machian
kinetic energy between two spinning bodies. 
\item [{{c)}}] The notion that the speed of light is not a postulate
but inherent in Machian inertia. 
\item [{{d)}}] Newtonian kinetic energy in the CM frame (only) reflects
the energy relationship between an object and the distant masses of
the universe, while the (anisotropic) Machian energy terms add the
energy relations between the various objects in the system under study
(e.g. Mercury and the sun). These small Machian terms account for
relativistic trajectories, like the anomalous precession of Mercury
and Lense-Thirring frame dragging. The Machian model applies to arbitrary
configurations of N-body systems. 
\item [{{e)}}] The notion that inertia is anisotropic. Isotropic models
can not explain relativistic trajectories. 
\item [{{f)}}] Relativistic effects of remote observation, such as time
dilation and redshift, require an isotropic model. 
\item [{{g)}}] The relational principle (including Mach's principle),
by which space, time and inertia emerge from the distribution of matter.
Simple models are proposed to relate these quantities to gravitational
potential, leading to an isotropic relational metric for general spacetimes,
which accounts for relativistic effects of observation equivalently
to GR. 
\item [{{h)}}] The relational model, i.e. the anisotropic Machian model
subject to the isotropic relational metric, accounts for both relativistic
trajectories and relativistic effects of observation equivalently
to GR. 
\item [{{i)}}] The notion that, from the relational point of view, the
GR-vacuum is not empty, but a space with a flat background potential
$\varphi_{o}$. 
\item [{{j)}}] From the relational point of view, the Schwarzschild dilation
factor agrees with a specific case of the more general relational
dilation factor. The latter possibly provides a mechanism to explain
phenomena at the galactic and cosmic scale, such as the flat galaxy
rotation curves and the accelerated expansion of the universe. 
\end{lyxlist}
It is not necessarily time dilation that makes a clock run slower.
So even fundamental concepts may be exchangeable, allowing for different
views of the same world. This makes the relational approach viable,
both in its own right and for its consistency with GR. I maintain
some reservations though, reminiscent of John Wheeler%
\footnote{\textquotedbl{}Spacetime tells matter how to move; matter tells spacetime
how to curve.\textquotedbl{}%
}, about the self-referential definitions of gravitational potential
and distance; one determines the other and vice versa.

\section*{Appendix A - Isotropic inertia can not explain anomalous perihelion
precession}

We represent the isotropic partial inertia between the planet $m$
and the sun $M$ by an arbitrary function $\xi=\xi(r)$, where $r$
is the separation between $m$ and $M$. Instead of the anisotropic
energy equation (\ref{eq:E orbit merc}), we have the isotropic equation
\begin{equation}
{\scriptstyle \frac{1}{2}}(m+\xi)\dot{r}^{2}+{\scriptstyle \frac{1}{2}}(m+\xi)r^{2}\dot{\phi}^{2}+(m+\mu)\varphi_{o}=E_{o}\label{eq:E merc iso}
\end{equation}
(where $\mu\varphi_{o}$ is the potential energy between planet and
sun, as in the anisotropic case). Note that the isotropic equation
is nearly the Newtonian equation
\begin{equation}
{\scriptstyle \frac{1}{2}}m\dot{r}^{2}+{\scriptstyle \frac{1}{2}}mr^{2}\dot{\phi}^{2}+(m+\mu)\varphi_{o}=E_{o}.\label{eq:E merc Newton}
\end{equation}
The isotropic inertia $\xi$ has an effect on the velocities $\dot{r}$
and $\dot{\phi}$. But this effect is equal in both directions, therefore
does not affect the trajectory $r(\phi)$ of the planet in the $r,\phi$-plane.
The easy way to show this is by replacing in (\ref{eq:E merc iso})
the time parameter $\tau$ by $\tau'=\tau m/(m+\xi)$, yielding the
Newtonian equation. Hence, the trajectory of the isotropic system
is Newtonian, therefore, is elliptic and has zero precession.

\section*{Appendix B - Basic Machian kinematic equation for two bodies}

We shall adopt a Lagrangian approach in deriving the equations of
motion for a system of two point particles $m_{1}$ and $m_{2}$ falling
to each other. This is done for two different cases: in empty space
and in our universe:

\subsection*{In empty space ($\boldsymbol{\varphi_{b}=0}$) }

The Lagrangian associated with the energy of the particles is
\begin{equation}
L=T-V\label{eq:Mach two-body empty energy}
\end{equation}
where
\begin{equation}
T={\scriptstyle \frac{1}{2}}\mu\dot{r}^{2}\qquad V=\mu\varphi_{o}.\label{eq:T and V 2-body empty}
\end{equation}
The inertia between the two bodies is $\mu=-Gm_{1}m_{2}/r\varphi_{o}$
(the subscripts of $\mu$ and $r$ have been dropped for convenience,
so $\mu=\mu_{12}$ and $r=r_{12}$). Lagrange's equation for $r$
is
\begin{equation}
\frac{d}{dt}\Bigl(\frac{\partial L}{\partial\dot{r}}\Bigr)-\frac{\partial L}{\partial r}=\dot{\mu}\dot{r}+\mu\ddot{r}+\frac{\mu}{r}({\scriptstyle \frac{1}{2}}\dot{r}^{2}-\varphi_{o})=0,
\end{equation}
which yields, using $\dot{\mu}=-\frac{\mu}{r}\dot{r}$, the Machian
kinematic equation for two masses falling to each other in empty space:
\begin{equation}
\ddot{r}=\frac{\varphi_{o}}{r}({\scriptstyle {\textstyle 1}}+\frac{\dot{r}^{2}}{{\textstyle {\textstyle {\textstyle {\scriptstyle 2}}}}\varphi_{o}}).\label{eq:kinematic eq empty}
\end{equation}
This differential equation lets $m_{1}$ and $m_{2}$ accelerate towards
each other until $\ddot{r}=0$, at which point the relative velocity
$|\dot{r}|$ attains its maximum value
\begin{equation}
|\dot{r}|_{max}=\sqrt{{\scriptstyle -}2\varphi_{o}}.\label{eq:speed limit}
\end{equation}
According to (\ref{eq:phi o}), $\varphi_{o}={\scriptstyle -}{\scriptstyle \frac{1}{2}}c^{2}$,
hence, $|\dot{r}|_{max}=c$, the speed of light. This is not a surprise,
as we recognize the Lorentz factor $\gamma$ in Eq.(\ref{eq:kinematic eq empty}):
\begin{equation}
({\scriptstyle {\textstyle 1}}+\frac{\dot{r}^{2}}{{\textstyle {\textstyle {\textstyle {\scriptstyle 2}}}}\varphi_{o}})^{{\scriptscriptstyle -1/2}}=({\scriptstyle {\textstyle 1}}{\scriptstyle -}\frac{\dot{r}^{2}}{c^{2}})^{{\scriptscriptstyle -1/2}}=\gamma.\label{eq:Lorentz factor}
\end{equation}
Fortunately, differential equation (\ref{eq:kinematic eq empty})
is easy to solve
\begin{equation}
r(t)=\biggl(\frac{\dot{r}_{0}^{2}{\scriptstyle -}c^{2}}{{\scriptstyle {\textstyle 4}}r_{0}}\biggr)\, t^{2}+\dot{r}_{0}t+r_{0}\label{eq:r(t)}
\end{equation}
where $r_{0}$ are $\dot{r}_{0}$ are initial values of $r$ and $\dot{r}$,
respectively. If we set $\dot{r}_{0}=0$, then (\ref{eq:r(t)}) reduces
to
\begin{equation}
r(t)=\frac{{\scriptstyle -}c^{2}}{{\scriptstyle {\textstyle 4}}r_{0}}\, t^{2}+r_{0}\label{eq:r(t) v(0)=00003D00003D0}
\end{equation}
so,
\begin{equation}
\dot{r}(t)=\frac{{\scriptstyle -}c^{2}}{2r_{0}}t\label{eq:r.}
\end{equation}
from which follows that collision ($r=0$) takes place at $t_{c}=2r_{0}/c$,
at which point
\begin{equation}
|\dot{r}(t_{c})|=|\frac{{\scriptstyle -}c^{2}}{2r_{0}}t_{c}|=c
\end{equation}

\subsection*{In our universe ($\boldsymbol{\varphi_{b}=\varphi_{o}}$). }

If we move $m_{1}$ and $m_{2}$ into our universe, somewhere near
earth, where $\varphi_{b}=\varphi_{o}$, then the kinetic energy term
in (\ref{eq:Mach two-body empty energy}) extends to
\begin{equation}
T={\scriptstyle \frac{1}{2}}\mu\dot{r}^{2}+{\scriptstyle \frac{1}{2}}m_{1}\dot{r}_{1}^{2}+{\scriptstyle \frac{1}{2}}m_{2}\dot{r}_{2}^{2}\label{eq:Mach two body universe energy}
\end{equation}
where $\dot{r}=\dot{r}_{1}+\dot{r}_{2}$ and where $\dot{r}_{1}$
and $\dot{r}_{2}$ are the velocities of the bodies relative to their
CM. We may simplify (\ref{eq:Mach two body universe energy}) to
\begin{equation}
T={\scriptstyle \frac{1}{2}}(\mu+m_{12})\dot{r}^{2}
\end{equation}
where $m_{12}=\frac{m_{1}m_{2}}{m_{1}+m_{2}}$ represents the additional
inertia due to the presence of the masses of the universe. Via the
Lagrangian, we obtain the kinematic equation
\begin{equation}
\ddot{r}=\frac{\mu}{\mu+m_{12}}\frac{\varphi_{o}}{r}({\scriptstyle {\textstyle 1}}+\frac{\dot{r}^{2}}{{\textstyle {\textstyle {\textstyle {\scriptstyle 2}}}}\varphi_{o}})\label{eq:kinematic eq universe}
\end{equation}
which is identical to the equation for the empty space case (\ref{eq:kinematic eq empty}),
except for an additional 'attenuation' factor $\lambda=\frac{\mu}{\mu+m_{12}}$.
Note that in general $\lambda\ll1$. As a result, the acceleration
of the particles toward each other is strongly attenuated by the extra
inertia $m_{12}$ caused by the cosmic masses. However, in the limit
for $r{\scriptstyle \rightarrow}0$, $\mu$ grows to infinity and
the attenuation disappears ($\lambda{\scriptstyle \rightarrow}1$),
thus the kinematic equation (\ref{eq:kinematic eq universe}) is asymptotically
equal to the kinematic equation (\ref{eq:kinematic eq empty}) for
the empty space case. So finally $\dot{r}$ will still reach the speed
of light at collision, in a much steeper acceleration though, as can
be verified easily by numerical simulation.\medskip{}

\noindent Hence, at collision, $m_{1}$ and $m_{2}$ reach the speed
of light towards each other, whatever the initial conditions are and
with or without the cosmic masses present. The condition for this
to happen, however, is that the distance between the point particles
$m_{1}$ and $m_{2}$ must actually get to zero. The question is how
realistic such collisions are, considering that particles have finite
size and other forces are involved in particle collisions. The mechanism
of (\ref{eq:kinematic eq universe}) suggests that the $r=0$ condition
may be (almost) met by particles which can get (almost) at the velocity
limit $c$, like photons and neutrino's. Another question is what
happens during and after collision? The extraordinary property of
the Machian kinematic equation (\ref{eq:kinematic eq universe}) is
that the state of the particles at collision ($r=0$, $\dot{r}=c$,
$\ddot{r}=0$) is always the same, i.e. does \textit{not} depend on
the state of the particles at any instance before collision. This
implies that also after collision the state of the particles does
not depend on pre-collision conditions (meaning history is gone).
From particle physics, however, we know that momentum and energy are
being conserved in both annihilation and pair production, i.e., state
history \textit{is} conserved in these collisions. This suggests that
the $r=0$ condition is not fully met in such collisions. This, in
turn, suggests that the particles have not quite reached the velocity
limit $c$.

\end{document}